\documentclass[twocolumn,trackchanges]{aastex701}
\usepackage{makecell}
\usepackage{multirow}
\usepackage{comment}
\usepackage{booktabs}
\usepackage{tabularx}
\usepackage[utf8]{inputenc}
\usepackage{newunicodechar}
\newunicodechar{，}{,}
\usepackage{array}
\usepackage{color}

\begin{document}

%\title{Correlated Flux Variations Between Fast Radio Bursts and their Associated Persistent Radio Emission: Insights into Shared Progenitors}
\title{Flux Variations of Fast Radio Bursts and Their Persistent Radio Sources: Evidence for a Shared Progenitor}

\author[orcid=0009-0008-0639-3964]{Xinming Li}
\affiliation{Laboratory for Compact Object Astrophysics and AstroSpace Technology, Central China Normal University, Wuhan 430079, China}
\affiliation{Institute of Astrophysics, Central China Normal University, Wuhan 430079, China}
\email{lixinming@mails.ccnu.edu.cn}

\author[orcid=0000-0001-6651-7799]{Chen-Hui Niu}
\affiliation{Laboratory for Compact Object Astrophysics and AstroSpace Technology, Central China Normal University, Wuhan 430079, China}
\affiliation{Institute of Astrophysics, Central China Normal University, Wuhan 430079, China}
\email[show]{niuchenhui@ccnu.edu.cn}

\author[orcid=0009-0009-3949-4726]{Jia-heng Zhang} 
\affiliation{Laboratory for Compact Object Astrophysics and AstroSpace Technology, Central China Normal University, Wuhan 430079, China}
\affiliation{Institute of Astrophysics, Central China Normal University, Wuhan 430079, China}
\email{jiahengzhang@mails.ccnu.edu.cn}

\author[orcid=0000-0003-3010-7661] {Di Li}
\affiliation{New Cornerstone Science Laboratory，Department of Astronomy，Tsinghua University，Beijing 100084, China}
\affiliation{National Astronomical Observatories, Chinese Academy of Sciences, Beijing 100101, China}
\email{dili@tsinghua.edu.cn}

\author[orcid=0000-0002-9725-2524] {Bing Zhang}
\affiliation{The Hong Kong Institute for Astronomy and Astrophysics, The University of Hong Kong,
Pokfulam, Hong Kong, P. R. China}
\affiliation{Department of Physics, The University of Hong Kong, Pokfulam, Hong Kong, P. R. China}
\affiliation{Nevada Center for Astrophysics, University of Nevada, Las Vegas, USA}
\affiliation{Department of Physics and Astronomy, University of Nevada, Las Vegas, USA}
\email{bzhang1@hku.hk}

\author[orcid=0000-0001-6374-8313] {Yuan-Pei Yang}
\affiliation{South-Western Institute for Astronomy Research, Yunnan Key Laboratory of Survey Science,
Yunnan University, Kunming, P. R. China}
\email{ypyang@ynu.edu.cn}

\author[orcid=0000-0001-9036-8543] {Pei Wang}
\affiliation{National Astronomical Observatories, Chinese Academy of Sciences, Beijing 100101, China}
\email{wangpei@nao.cas.cn}

\author[orcid=0009-0005-8586-3001] {Junshuo Zhang}
\affiliation{National Astronomical Observatories, Chinese Academy of Sciences, Beijing 100101, China}
\affiliation{School of Astronomy and Space Science, University of Chinese Academy of Sciences, Beijing 100049, China}
\email{zhangjs@bao.ac.cn}

\author[orcid=0000-0002-8744-3546] {Yongkun Zhang}
\affiliation{National Astronomical Observatories, Chinese Academy of Sciences, Beijing 100101, China}
\email{zhangyongkun15@mails.ucas.ac.cn}

\author[orcid=0000-0001-5931-2381] {Ye Li}
\affiliation{Purple Mountain Observatory, Chinese Academy of Sciences, Nanjing, P. R. China}
\affiliation{School of Astronomy and Space Sciences, University of Science and Technology of China,
Hefei, P. R. China}
\email{yeli@pmo.ac.cn}

\author[orcid=0000-0001-8065-4191] {Jiarui Niu}
\affiliation{National Astronomical Observatories, Chinese Academy of Sciences, Beijing 100101, China}
\email{niujiarui@nao.cas.cn}

\author[orcid=0000-0001-8868-4619] {Xiaoping Zheng}
\affiliation{Laboratory for Compact Object Astrophysics and AstroSpace Technology, Central China Normal University, Wuhan 430079, China}
\affiliation{Institute of Astrophysics, Central China Normal University, Wuhan 430079, China}
\email{zhxp@ccnu.edu.cn}

\author[orcid=0000-0002-1067-1911] {Yunwei Yu}
\affiliation{Laboratory for Compact Object Astrophysics and AstroSpace Technology, Central China Normal University, Wuhan 430079, China}
\affiliation{Institute of Astrophysics, Central China Normal University, Wuhan 430079, China}
\email{yuyw@ccnu.edu.cn}

\author[orcid=0000-0002-0475-7479] {Yi Feng}
\affiliation{Research Center for Astronomical Computing, Zhejiang Laboratory, Hangzhou, P. R. China}
\affiliation{Institute for Astronomy, School of Physics, Zhejiang University, Hangzhou, P. R. China}
\email{yifeng@nao.cas.cn}

\author[orcid=0000-0003-1353-9040]{Bo Zhang}
\affiliation{Shanghai Astronomical Observatory, 80 Nandan Road, Shanghai 200030, People’s Republic of China}
\email{zb@shao.ac.cn}

\author[orcid=0000-0003-4157-7714] {Fayin Wang}
\affiliation{School of Astronomy and Space Science, Nanjing University, Nanjing 210093, China}
\affiliation{Key Laboratory of Modern Astronomy and Astrophysics (Nanjing University), Ministry of Education, Nanjing 210093, China}
\email{fayinwang@nju.edu.cn}

\author[orcid=0009-0009-8320-1484] {Yuhao Zhu}
\affiliation{National Astronomical Observatories, Chinese Academy of Sciences, Beijing 100101, China}
\affiliation{School of Astronomy and Space Science, University of Chinese Academy of Sciences, Beijing 100049, China}
\email{zhuyh@bao.ac.cn}

\author[orcid=0000-0003-1838-8456] {A.Ming Chen}
\affiliation{Laboratory for Compact Object Astrophysics and AstroSpace Technology, Central China Normal University, Wuhan 430079, China}
\affiliation{Institute of Astrophysics, Central China Normal University, Wuhan 430079, China}
\email{chensm@mails.ccnu.edu.cn}

\author[orcid=0000-0003-3572-2364] {Ze-Xin Du}
\affiliation{Laboratory for Compact Object Astrophysics and AstroSpace Technology, Central China Normal University, Wuhan 430079, China}
\affiliation{Institute of Astrophysics, Central China Normal University, Wuhan 430079, China}
\email{duzexin@mails.ccnu.edu.cn}

\author[orcid=0000-0003-1720-9727]{Jian Li}
\affiliation{Department of Astronomy, University of Science and Technology of China, Hefei, China}
\affiliation{School of Astronomy and Space Sciences, University of Science and Technology of China,
Hefei, P. R. China}
\email{jianli@ustc.edu.cn}

\author[orcid=0009-0009-4317-5920]{weihong li}
\affiliation{Laboratory for Compact Object Astrophysics and AstroSpace Technology, Central China Normal University, Wuhan 430079, China}
\affiliation{Institute of Astrophysics, Central China Normal University, Wuhan 430079, China}
\email{liwwww0605@163.com}

\author[orcid=0000-0002-9441-2190]{Chenchen Miao}
\affiliation{Department of Astronomy, College of Physics and Electronic Engineering, Qilu Normal University, Jinan 250200, China}
\affiliation{Shandong Key Laboratory of Space Environment and Exploration Technology, No. 180, Wenhua West Road, Weihai City, Shandong Province 264209, People's Republic of China}
\email{miaocc@qlnu.edu.cn}

\author[orcid=0000-0001-9036-8543] {Weiyang Wang}
\affiliation{School of Astronomy and Space Science, University of Chinese Academy of Sciences, Beijing 100049, China}
\email{wywang@ucas.ac.cn}

\author[orcid=0000-0001-8744-3813] {Guang-Lei Wu}
\affiliation{Laboratory for Compact Object Astrophysics and AstroSpace Technology, Central China Normal University, Wuhan 430079, China}
\affiliation{Institute of Astrophysics, Central China Normal University, Wuhan 430079, China}
\email{wuguanglei@mails.ccnu.edu.cn}

\author[orcid=0000-0003-4546-2623] {Ai Yuan Yang}
\affiliation{National Astronomical Observatories, Chinese Academy of Sciences, Beijing 100101, China}
\affiliation{Key Laboratory of Radio Astronomy and Technology, Chinese Academy of Sciences, A20 Datun Road, Chaoyang District, Beijing, 100101, P. R. China}
\email{yangay@bao.ac.cn}

\author[orcid=0000-0002-4997-045X]{Jumei Yao}
\affiliation{Xinjiang Astronomical Observatory, Chinese Academy of Sciences, Urumqi 830011, China}
\email{yaojumei@xao.ac.cn}

\author[orcid=0000-0002-1243-0476]{Ru-Shuang Zhao}
\affiliation{Guizhou Normal University, Guiyang 550001, China}
\email{201907007@gznu.edu.cn}

\begin{abstract}
Fast radio bursts (FRBs) are millisecond-duration extragalactic radio transients, some of which are associated with compact persistent radio sources (PRSs), hinting at a physical connection. While several models have been proposed to explain PRSs and their connection to FRBs, direct observational tests remain limited. Here, we report for the first time a correlated trend between the long-term variation of the PRS flux density and the burst energetics of FRB 20190520B and FRB 20240114A, suggesting a physical coupling between the PRS and FRB activity. We further examine additional repeaters with compact PRSs and find no clear correlation between PRS luminosity and burst activity, likely due to the limited observations. These results are consistent with scenarios in which both the PRS and FRB activity may be powered by a common energy reservoir, such as the magnetic or rotational energy of a magnetar.
\end{abstract}


\section{INTRODUCTION}
Fast radio bursts (FRBs) are millisecond-duration radio transients with extremely high brightness temperatures at extragalactic distances, whose central engine and emission physics remain debated \citep{2007Sci...318..777L,2021ApJS..257...59C,2022A&ARv..30....2P,RevModPhys.95.035005}. As the detected sample has grown, a subset of FRBs has been found to repeat \citep{2016Natur.531..202S}. The discovery of repeating activity is pivotal, enabling precise localizations and systematic follow-up across multiple epochs and frequencies \citep{2017Natur.541...58C,2017ApJ...834L...8M}. 

For several well-localized repeating FRBs, compact, non-thermal persistent radio sources (PRSs) have been found in spatial coincidence with the burst positions. These sources are typically unresolved on milliarcsecond scales and exhibit GHz-band power-law spectra, with radio luminosities reaching $L_\nu \sim 10^{27\text{–}29}\,{\rm erg\,s^{-1}\,Hz^{-1}}$. Typical cases include two PRSs spatially associated with FRB 20121102A \citep{2017Natur.541...58C,2017ApJ...834L...8M} and FRB 20190520B \citep{2022Natur.606..873N}, along with lower-luminosity PRSs associated with FRB 20201124A \citep{2022MNRAS.513..982R,2024Natur.632.1014B}, FRB 20240114A \citep{2025AA...695L..12B,2025arXiv250114247Z} and FRB 20190417A \citep{2025arXiv250905174M}. At the same time, many repeating FRBs show no detectable PRS, even at deep observational limits \citep{2020Natur.577..190M,2021ApJ...919L..24B}. This indicates that bright PRSs may not be a universal feature of such repeaters, or that their sources may lie at different evolutionary stages.

Whether PRSs are physically linked to FRBs remains a central question. An empirical correlation between PRS luminosity and $\left|\rm{RM}\right|$, where RM is an FRB observable characterizing the Faraday rotation of the polarized burst signal, has been proposed \citep{2020ApJ...895....7Y}. The most luminous PRS systems tend to show extreme RM ($|{\rm RM}|\sim 10^{4\text{–}5}\,{\rm rad\,m^{-2}}$), while lower-luminosity PRSs exhibit milder RMs $(\sim 10^2\,{\rm rad\,m^{-2}})$. On intra-source timescales, short-term correlations between PRS flux variations and FRB burst energy are often insignificant. Some sources show modulation beyond what is expected from pure refractive interstellar scintillation (RISS), indicating intrinsic PRS variability or a complex scattering environment \citep{2017Natur.541...58C,2025arXiv250114247Z}.

Theoretically, a young magnetar is one of the most widely discussed explanations for FRBs and their associated persistent radio emission. Even prior to the discovery of the first PRS, it was proposed that the relativistic outflow from a young magnetar could inflate a magnetar wind nebula (MWN), producing long-lived synchrotron emission \citep{2016ApJ...819L..12Y}. In this framework, the relativistic wind launched by the magnetar inflates a nebula that powers the persistent radio emission through synchrotron radiation from relativistic electrons, naturally linking high $\left|\rm{RM}\right|$, large host-galaxy dispersion measures, and a dense, turbulent local environment \citep{2017ApJ...841...14M,2018ApJ...868L...4M}.
This scenario is closely related to the earlier pulsar wind nebula (PWN) paradigm and was later applied to explain the compact PRS associated with FRB 20121102A. Interaction between the MWN and the surrounding supernova remnant can naturally produce a composite system in which the nebula powers the persistent radio emission while the expanding remnant dominates the propagation effects, such as large and evolving RM and dispersion measure (DM) \citep{2021ApJ...923L..17Z}. Additionally, alternative classes of models invoke accretion-powered compact objects. In the “hypernebula” scenario, powerful outflows launched by a hyper-accreting compact object inflate a compact synchrotron nebula and can account for the observed PRS luminosity and large Faraday rotation \citep{2022ApJ...937....5S}. Other accretion-powered scenarios associate the PRS with radio emission from low-luminosity active galactic nuclei (AGNs) or massive/intermediate-mass black holes in dense galactic environments \citep{2017A&A...602A..64V,Katz_2017}. These models suggest that PRSs may be powered by distinct compact objects or energy sources, requiring future observations to determine the dominant mechanisms behind these emissions.

Among repeating FRBs associated with PRSs, FRB 20190520B and FRB 20240114A stand out for their persistent activity and exceptionally high burst rates, respectively. FRB 20190520B sustained multi-year activity and extreme, rapidly evolving propagation properties. Its DM is exceptionally large and declining steadily at $\sim 10\ {\rm pc\ cm^{-3}\ yr^{-1}}$ \citep{NIU202676}. Additionally, its RM reaches extreme absolute values and undergoes dramatic temporal evolution, including sign reversals. Highly variable millisecond-scale scattering further indicates a dynamic, inhomogeneous local plasma environment \citep{2023MNRAS.519..821O}. These properties suggest a young magnetar embedded in a dense and highly magnetized environment, possibly associated with an expanding supernova remnant \citep{2025arXiv251207140W}, where the compact PRS naturally arises from a MWN. Furthermore, its predominantly narrowband, intrinsically generated bursts point to a magnetospheric origin \citep{2024ChPhL..41j9501Z,zhang2025statisticaltemporalanalysismulticomponent}. FRB 20240114A was first reported by CHIME \citep{2024ATel16420....1S}. Subsequent follow-up observations with Five-hundred-meter Aperture Spherical radio Telescope (FAST) revealed an exceptionally high burst rate, with more than 11,000 bursts detected over a monitoring window of approximately 214 days, making it one of the most burst-active FRB sources identified to date \citep{2025arXiv250714707Z}. Despite this intense bursting activity, the RM has not exhibited significant temporal variation over the limited epochs currently available \citep{2025ApJS..278...49X}. 

Despite these advances, the connection between PRS properties and FRB activity remains poorly understood. Assuming a co-source magnetar model, we expect these components to be physically coupled within the system. Specifically, progenitor energy releases and FRB bursts may share a common physical origin, or energy from the FRBs could be deposited into the surrounding nebula, driving contemporaneous or delayed variations in the PRS flux density. These expectations motivate our systematic investigation of the correlation between PRS emission and FRB energy output.

In this paper, we investigate the statistical and physical connections between repeating FRBs and their associated PRSs. Section 2 describes the compiled dataset, introduces a proxy for the central engine’s daily energy output, and outlines the methods used to analyze its correlation with PRS emission.
Section 3 presents the observational results and the quantitative outcomes of the correlation analysis.
Section 4 discusses the physical implications of these findings, including the intrinsic origin of PRS variability and the possibility of a shared energy reservoir that powers both FRB bursts and PRS emission.
Finally, Section 5 summarizes our main conclusions.

%% The "ht!" tells LaTeX to put the figure "here" first, at the "top" next
%% and to override the normal way of calculating a float position.
%% The asterisk after "figure" tells the compiler to span multiple columns
%% if a two column style is selected.

\section{data and analysis methods} \label{data sets and analysis methods}
\begin{table*}[ht]
\centering
\footnotesize
\caption{L-band FRB Observations and PRS Flux Density Measurements for Five FRB–PRS Systems}
\label{tab:frb_prs}
\begin{tabular}{l @{\hspace{0.8em}} p{3.6cm} p{2.6cm} p{2cm} @{\hspace{0.8em}} p{3.4cm} l l}
\hline

% ===== 控制行距（关键）=====
%[-0.3em]  % 可以改成 [-0.2em] 或 [-0.5em] 微调

% ===== 第二行（原表头）=====
Source & Ref(Flux density,PRS) & Observation time$^a$ & telescope$^b$  &
Ref(Burst rate,FRB) & Observation time$^c$ & telescope$^d$ \\
\hline
% ---------- FRB 20121102A ----------
\multirow{8}{*}{20121102A}
  & \cite{2017Natur.541...58C} & 2016.4--9 (37)      & VLA(S)    
  & \cite{2022MNRAS.515.3577H}      & 2015.11--2016.10 & Arecibo \\
  & \cite{2017ApJ...834L...8M}     & 2016.2--9 (6)       & EVN(L,C)
   &\cite{2021Natur.598..267L}          & 2019.8--10     & FAST\\
  & \cite{2021AA...655A.102R}       & 2017.5--12 (4)      & GMRT(P,L)
  &\cite{2025arXiv250715790W}         &2020.3--2023.4    &FAST\\
  & \cite{2022MNRAS.511.6033P}      & 2017.2--11 (6)      & EVN(L,C)\\
  & \cite{2023ApJ...958..185C}        & 2017.5--8 (7)       & VLA(Ku,K)\\
  &\cite{2023MNRAS.525.3626R}      & 2019.9--2022.9 (5)  & MeerKAT(L) \\
  &\cite{2024ApJ...976..165Y}        & 2023.5--6 (9)       & VLA(L) \\
  & \cite{bhardwaj2025constrainingoriginfrb20121102as}    & 2022.11--2023.8 (8) & GMRT(P,L) \\
\hline
% ---------- FRB 20190520B ----------
\multirow{5}{*}{20190520B}
  & \cite{2022Natur.606..873N}               & 2020.7--11 (14)        & VLA(L,S,C)
  & \cite{2022Natur.606..873N}                & 2020.4--2020.9        & FAST \\
  & \cite{2023ApJ...959...89Z}             & 2021.10--11 (8)        & VLA(L,S,C,X)& \cite{NIU202676}.               & 2020.4--2023.3        & FAST \\
  &\cite{2023ApJ...958L..19B}          & 2022.2 (2)             & EVN(L) & In prep. (Zhang)                & 2020.4--2024.12        & FAST \\
  & \cite{2024ApJ...976..165Y}             & 2023.6 (7)             & VLA(S)\\
  & \cite{Balasubramanian_2025}   & 2022.11--2024.9 (8)    & GMRT(L)\\
\hline
% ---------- FRB 20201124A ----------
\multirow{2}{*}{20201124A}
  & \cite{2024Natur.632.1014B}               & 2021.10--2022.3 (3)        & VLA(C,Ku,K)
  & \cite{2022Natur.609..685X}               & 2021.4--6        & FAST \\
  &- &- &- &\cite{2022RAA....22l4002Z}    & 2021.9-10       & FAST\\
\hline
% ---------- FRB 20240114A ----------
\multirow{3}{*}{20240114A}
  &\cite{2025AA...695L..12B}            & 2024.9 (2)        & VLBA(C) 
  &\cite{2025arXiv250714707Z}                & 2024.1--2024.8        & FAST \\
  & \cite{2025arXiv250114247Z}              & 2024.2--10 (11)        & VLA(L,S,C) MeerKAT(L)\\
  & \cite{bhusare2025lowfrequencyprobespersistentradio}          & 2024.2--8 (7)             & GMRT(P)\\
\hline
% ---------- FRB 20190417A ----------
\multirow{1}{*}{20190417A}
  &\cite{2025arXiv250905174M}              & 2021.10--2022.8(1)       & EVN(L)
   & \cite{2025SCPMA..6889511F}               & 2020.8--2022.11        & FAST \\
\hline
\end{tabular}
\tablenotetext{a}{This column lists the time span of the PRS observations, with the number in parentheses indicating the total number of observing epochs within that interval.}
\tablenotetext{b}{This column specifies the telescope and the associated observing frequency band used in the PRS observations.}
\tablenotetext{c}{This column lists the time span of the FRB observations.}
\tablenotetext{d}{This column specifies the telescope used in the FRB observations.}
\end{table*}

\subsection{Data} \label{Data}
We compile data for five FRB–PRS systems that have been reported to host compact PRSs. For each system, we compile all available PRS flux-density measurements together with the FRB burst data obtained during the corresponding periods of PRS monitoring, focusing on observational campaigns that span extended time intervals. The burst sample is restricted to L-band observations, where the observational coverage is the most extensive for these repeaters, enabling a more uniform and statistically robust characterisation of their burst activity. The references for all PRS and burst data used in this work are summarised in Table \ref{tab:frb_prs}.
\subsection{A daily energy–output proxy}\label{subsec:energy_proxy}
To quantify the energy that each repeater injects into its surrounding environment on observable timescales, we construct a simple activity indicator based on the detected bursts. For each observing day, we calculate the total fluence of all detected bursts normalized by the effective observing time. This yields a daily energy–output proxy
\begin{equation}
P_{\rm day} \equiv \frac{\sum_{i=1}^{N} F_i}{T_{\mathrm{obs}}} \, ,
\end{equation}
where $N$ is the number of bursts detected on a given day, $F_i$ denotes the fluence of the $i$-th burst, and $T_{\rm obs}$ is the total effective observing time on that day. The quantity $P$ represents the burst fluence accumulated per unit observing time and serves as a proxy for the daily burst energy output.
\subsection{Correlation analysis method} \label{analysis methods}
\subsubsection{Time-window–based paired correlation analysis}
\label{method1}
We employ a time-window–based paired correlation method to test whether the FRB activity and the PRS emission exhibit synchronous variations on finite timescales, without assuming a fixed time lag. For each PRS observation epoch, we compute the mean FRB activity proxy $\langle P\rangle$ within a symmetric time window centered on that epoch, and pair it with the corresponding PRS flux density. Correlations between the paired quantities are evaluated using both Pearson and Spearman statistics.
\begin{figure*}[!ht]
    \centering
    \includegraphics[width=1\textwidth]{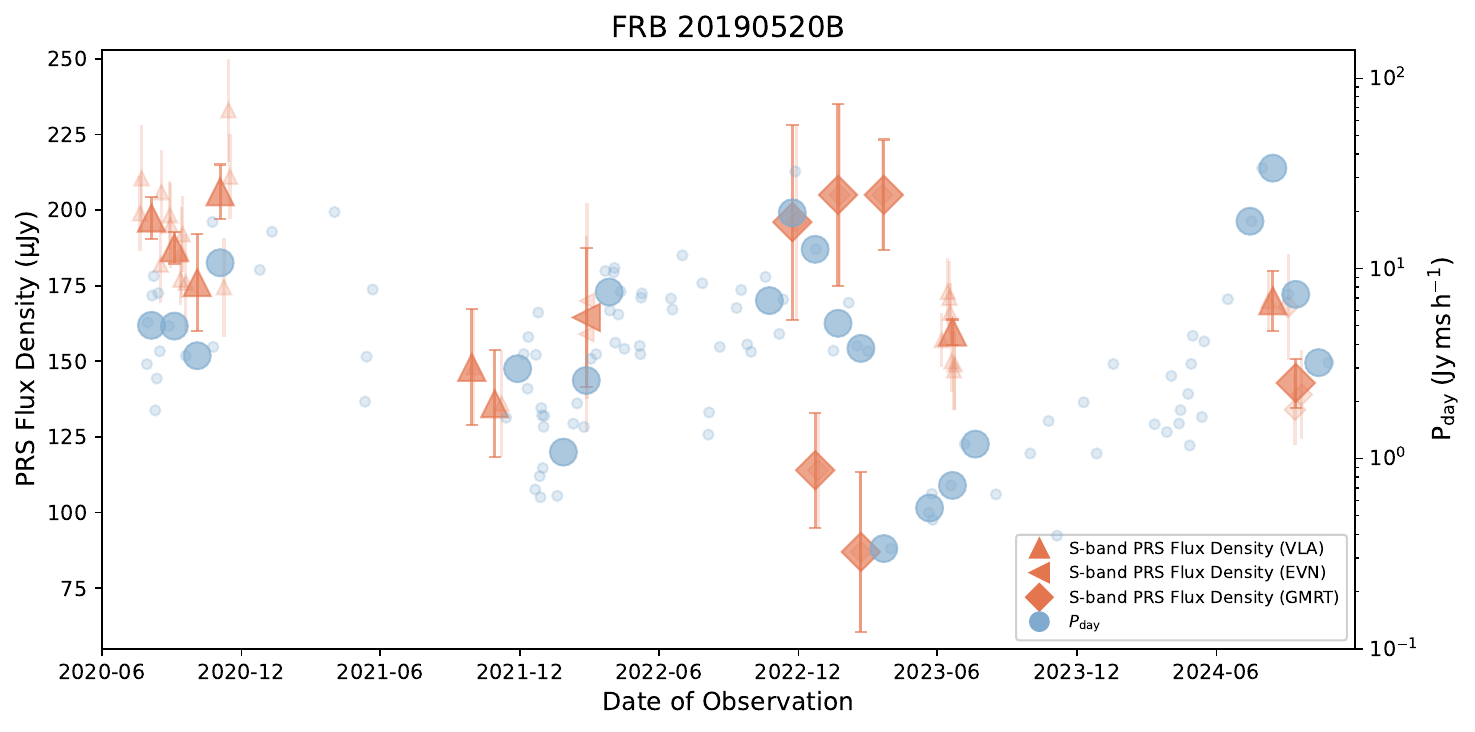}
    \caption{\textbf{PRS flux density of FRB 20190520B and the corresponding daily FRB activity proxy $P_{\rm day}$.} The smaller orange symbols denote individual PRS flux-density measurements, which have been uniformly converted to the S band under the assumption of a power-law spectrum. The smaller blue markers represent the daily energy-output proxy $P_{\rm day}$, calculated from FRB bursts detected by FAST. The detailed values of the S-band–converted PRS flux densities and the corresponding $P_{\rm day}$ are listed in Appendix Table \ref{tab:2} and Table \ref{tab:3}. The larger orange and blue symbols indicate the binned PRS flux density and $P_{\rm day}$, respectively, obtained using a 30-day binning scheme starting from the first PRS observation. In addition, the VLA observations obtained in June 2023 were analyzed in both \cite{2024ApJ...976..165Y} and \cite{Balasubramanian_2025}, yielding slightly different results due to differences in the data-reduction procedures. In this work, we adopt the most recent processing results presented in \cite{Balasubramanian_2025}.}
    \label{fig:1}
\end{figure*}

To assess statistical significance, we perform permutation tests by randomly shuffling the window-averaged FRB activity values while keeping the PRS flux measurements fixed, and recomputing the correlation coefficients. We adopt 5000 permutations in this analysis. This choice provides a stable estimate of the empirical $p$-values while maintaining reasonable computational efficiency, and is commonly used in permutation-based analyses. Because the windowed pairs are not strictly independent, the permutation-based $p$-values are adopted as the primary measure of significance. We adopt a null hypothesis that the PRS flux density and FRB burst activity are uncorrelated, and a threshold of $p < 0.05$ is used to identify statistically significant correlations.

We adopt representative time-window sizes comparable to the typical observational intervals of the data. These window sizes are not tuned to produce a specific outcome, but are chosen to explore the correlation behavior across different timescales. We find that the results remain qualitatively consistent across different window sizes, indicating that the conclusions are not sensitive to the exact binning scheme.

\subsubsection{Interpolated cross-correlation function (ICCF)}
Because the PRS and FRB observations are unevenly sampled and not strictly simultaneous, we employ the ICCF method \citep{1994PASP..106..879W} to investigate possible correlations between the PRS flux density and the daily energy-output proxy $P_{\rm day}$.

In the ICCF analysis, correlations are evaluated by shifting one time series relative to the other and linearly interpolating it onto the sampling epochs of the second series. To mitigate possible interpolation bias, the ICCF is computed in both interpolation directions, by alternatively interpolating $P_{\rm day}$ to the PRS observing epochs and interpolating the PRS flux densities to the epochs of $P_{\rm day}$. The final ICCF is then obtained by averaging the two resulting correlation functions.

We quantify the correlation using both the Pearson correlation coefficient, which measures linear relationships, and the Spearman rank correlation coefficient, which is sensitive to monotonic trends and less affected by outliers. Uncertainties are estimated using a flux-randomization bootstrap procedure \citep{Peterson_1998}, in which the PRS flux measurements are perturbed according to their reported uncertainties and resampled with replacement.

To evaluate the statistical significance of the correlation, we perform permutation tests by randomly shuffling the temporal order of one time series while keeping the other fixed. The ICCF is recomputed for each of 1000 permutations, and the empirical $p$-value is defined as the fraction of trials whose maximum correlation coefficient exceeds that obtained from the real data. We adopt the same null hypothesis of no intrinsic correlation between the two time series, and correlations with $p<0.05$ are considered statistically significant.

\section{analysis and results} \label{analysis and result}
In this section, we present a systematic investigation of the five currently available FRB–PRS systems based on the existing observational data. Sections \ref{a}, \ref{b}, and \ref{c} focus on FRB 20190520B, FRB 20240114A, and FRB 20121102A, respectively, as these three sources are more suitable for correlation analysis given the temporal coverage and sampling of their PRS and burst data. FRB 20201124A and FRB 20190417A are also included in the sample, but are not examined in the same level of detail because the available PRS measurements are too limited and the temporal overlap with the burst observations is insufficient for a reliable formal correlation analysis. For FRB 20201124A, Figure \ref{fig:7} shows the temporal distribution of its three PRS observations relative to the burst data. FRB 20190417A, by contrast, has only a single PRS measurement and is therefore not shown separately in a figure.
\subsection{FRB 20190520B}\label{a}

Within our sample, FRB 20190520B provides the most robust dataset. Its burst activity has been monitored exclusively with FAST over several years, thereby avoiding sensitivity differences among telescopes, and its PRS has been observed in multiple epochs since its discovery. 
To place all PRS flux-density measurements on a common frequency scale, we scale the reported values to 3 GHz assuming a power-law spectrum with an effective spectral index of $\alpha_\nu=-0.375$. This choice is based on Equation  
\begin{equation}
F(\nu,t)=A\,\nu^{-\left(\frac{\alpha-1}{2}\right)}
\left(\frac{t_{\mathrm{obs}}+t_{\mathrm{age}}}{t_{\mathrm{age}}}\right)^{-\left(\frac{\alpha^{2}+7\alpha-2}{4}\right)} \, .
\end{equation} 
and the corresponding fit results of \cite{Balasubramanian_2025}, who performed a joint modeling of the source using multi-frequency observations spanning 1.3–10 GHz within a magnetar wind nebula framework. In their formulation, $t_{\rm obs}$ denotes the time since the first observation, and $t_{\rm age}$ is the source age at that epoch. For measurements obtained at the same observing epoch, the time-dependent term cancels out, leaving the frequency dependence $F_\nu \propto \nu^{-(\alpha-1)/2}$. Substituting their best-fit $\alpha$ value gives an effective spectral index close to -0.375, which we therefore adopt here for the frequency conversion. This choice is also broadly consistent with the published GHz-band spectral slopes reported for this source. We further performed a robustness check using other reasonable spectral-index values, and found that neither the overall trend shown in Figure \ref{fig:1} nor the main qualitative conclusions changed.

\begin{table*}[ht]
\centering
\normalsize
\caption{Results of the time-window pairing analysis between the PRS flux density and the FRB burst-activity proxy for FRB 20190520B and FRB 20240114A. For each source, paired samples are constructed by averaging the burst-activity proxy within symmetric time windows of $\pm$ 5, $\pm$ 7.5, and $\pm$ 10 days centered on each PRS observing epoch. The table lists the Pearson correlation coefficient r, the Spearman rank coefficient $\rho$, and the corresponding permutation-based p-values (perm-p) derived from 5000 permutations.}
\label{tab:5}
\begin{tabular}{c c c c c c}
\hline\hline
source & Window & Pearson $r$ & perm-$p$ & Spearman $\rho$ & perm-$p$ \\
\hline
0520B & $\pm5$d   & +0.88 & $2.0\times10^{-4}$ & +0.86 & $2.0\times10^{-4}$ \\
      & $\pm7.5$d & +0.77 & $4.0\times10^{-4}$ & +0.79 & $4.0\times10^{-4}$ \\
      & $\pm10$d  & +0.12 & $0.6$             & +0.72 & $6.0\times10^{-4}$ \\
\hline
0114A & $\pm5$d   & +0.78 & $1.3\times10^{-2}$ & +0.84 & $8.0\times10^{-3}$ \\
      & $\pm7.5$d & +0.79 & $1.3\times10^{-2}$ & +0.87 & $3.8\times10^{-3}$ \\
      & $\pm10$d  & +0.82 & $8.4\times10^{-3}$ & +0.95 & $6.0\times10^{-4}$ \\
\hline
\end{tabular}
\end{table*}
The PRS light curve of FRB 20190520B shows noticeable variability over the observational baseline, as illustrated in Figure~\ref{fig:1}. In particular, the GMRT observations obtained around December 2022 display pronounced short-timescale fluctuations in the PRS flux density. The amplitude of these flux variations is significantly larger than that typically seen in other epochs , and therefore appears unusual. Such pronounced variability may be related to local environmental or propagation effects, such as absorption, scattering, or scintillation along the line of sight (LOS). This interpretation is supported by previous studies suggesting that FRB 20190520B resides in a dense and dynamically evolving local environment with significant scattering and propagation effects \citep{2022ApJ...931...87O}. These processes may temporarily modulate the observed PRS flux and perturb the underlying co-evolution between the PRS emission and FRB activity.

To mitigate the influence of these transient environmental effects, we restrict our correlation analysis to the remaining observational epochs. We then construct paired samples using the time-window pairing method described in Section \ref{method1} and evaluate their correlations using both Pearson and Spearman statistics. The statistical significance is assessed through 5000 permutation realizations. The results are summarized in Table \ref{tab:5}. Across different time-window configurations, the Spearman test consistently yields statistically significant positive correlations. The Pearson statistic, however, shows significant correlations at the $\pm5$ d and $\pm7.5$ d windows, but becomes insignificant at the $\pm10$ d window.
This difference is expected, as the Pearson coefficient measures strictly linear relationships and is more sensitive to scatter and outliers, whereas the Spearman statistic captures monotonic trends and is more robust to non-Gaussian variability. From a physical perspective, we do not necessarily expect a strictly linear relation between the PRS flux density and FRB activity; instead, a monotonic connection may arise if both are modulated by variations in the central engine. We nevertheless include the Pearson coefficient as a complementary diagnostic, providing a consistency check for linear trends and helping to assess the impact of scatter and outliers. The consistently significant Spearman correlations therefore indicate the presence of a robust monotonic relationship between the two quantities. These results suggest that the PRS flux density and FRB activity exhibit coordinated variability on weekly to monthly timescales.

\begin{figure*}[!ht]
    \centering
    \includegraphics[width=1\linewidth]{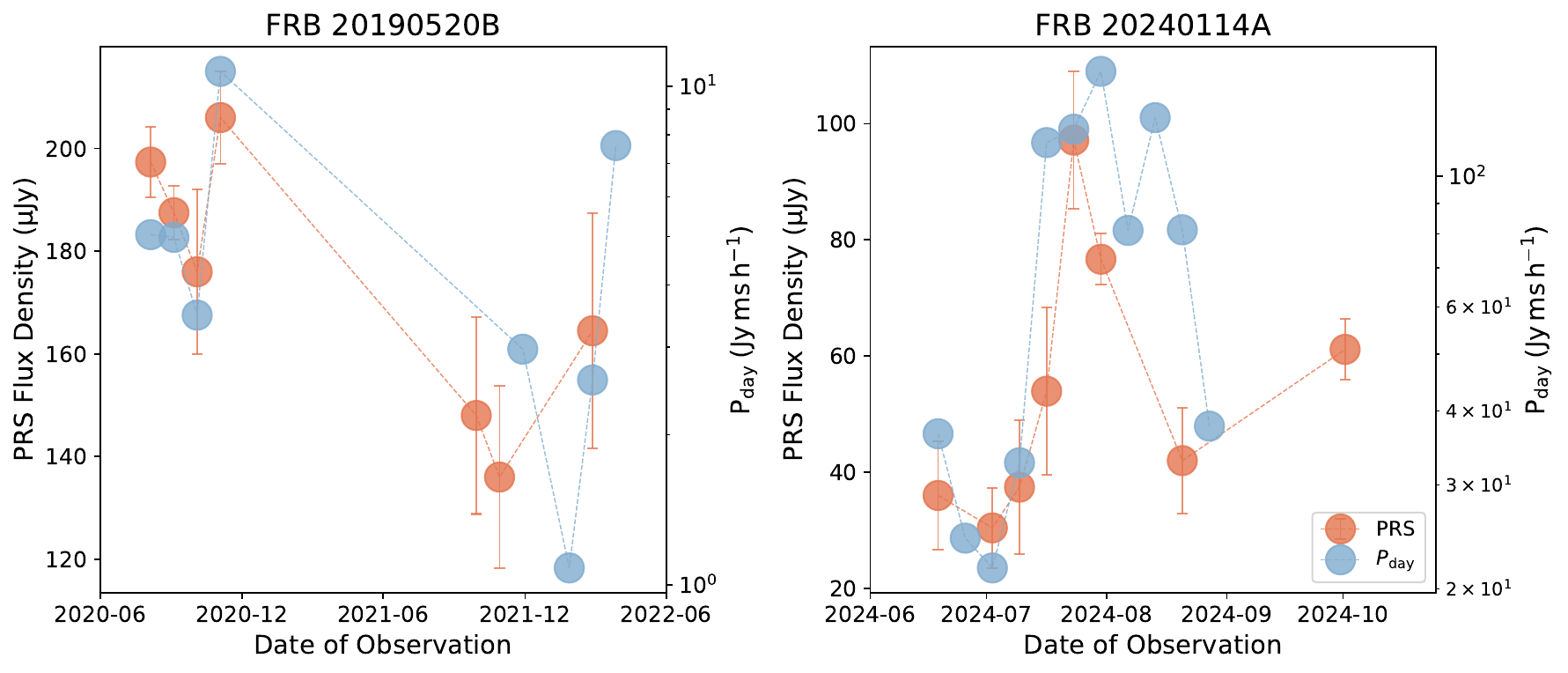}
    \caption{\textbf{Co-variation of PRS Flux Density and FRB Activity in FRB 20190520B and FRB 20240114A.} Both panels show binned results. A 30-day bin size (data from June 2020 to June 2022) is adopted for FRB 20190520B, while a 7-day binning scheme is used for FRB 20240114A. Because a single bin for FRB 20240114A may contain measurements from multiple telescopes, the binned PRS data are not further distinguished by telescope symbols.}
    \label{fig:placeholder}
\end{figure*}

To further investigate the temporal relationship between the two quantities, we also perform an ICCF analysis. However, during the later observational stage only a few PRS measurements are available, and these points are separated from the earlier data by a substantial temporal gap. Including such sparse measurements in the ICCF calculation would require extensive interpolation and could artificially enhance the apparent correlation. We therefore restrict the ICCF analysis to the earlier observing interval (from June 2020 to June 2022; see the left panel of Figure \ref{fig:placeholder}) where the PRS observations are relatively continuous.

Within this interval, the ICCF analysis shows that $P_{\rm day}$ and the PRS flux density remain positively correlated over a broad range of time lags. The correlation coefficient varies smoothly with lag and does not exhibit a sharp peak at any particular delay. In addition, the lag corresponding to the maximum correlation differs between correlation estimators, indicating that the time lag itself is not robustly constrained. Nevertheless, significance tests for the maximum correlation obtained with each estimator still indicate statistically meaningful correlations (see Appendix Figure~\ref{fig:2}).

Taken together, these results suggest the presence of a positive relationship between FRB activity and PRS emission in FRB 20190520B. Although the absence of a sharp correlation peak prevents a precise measurement of the time delay, the coherent variability observed on weekly to monthly timescales is consistent with a scenario in which both phenomena are driven by a common central engine. Variability on shorter timescales, however, may be influenced or partially masked by fluctuations in the local environment surrounding the source.

\subsection{FRB 20240114A}\label{b}
For FRB 20240114A, we homogenize all burst measurements obtained at different observing frequencies by converting them to an equivalent 3 GHz fluence, assuming a spectral index of $-0.34$ \citep{2025AA...695L..12B}, derived from VLBA observations over 1.3--5 GHz. Based on these rescaled fluences, we derive the corresponding burst activity proxy P for each observing epoch. The resulting time series is presented in Figure \ref{fig:placeholder}. Because the PRS and burst-activity measurements for this source are relatively well matched in time after binning, we first examine the correlation between the binned PRS flux density and the binned burst-activity proxy. We quantify the correlation using both the Pearson and Spearman coefficients, and assess the statistical significance through 5000 permutation realizations. This analysis reveals a clear positive correlation (Pearson $r$=+0.8352, perm-p=1.4$\times10^{-2}$; Spearman $\rho$=+0.9286, perm-p=7.399$\times10^{-3}$). To further assess the robustness of this finding, we apply the time-window–based pairing method described in Section \ref{method1}. The resulting correlation coefficients and permutation-based p-values are summarized in Table \ref{tab:5} and likewise indicate a positive correlation. 

To further investigate the temporal relationship between the two quantities, we perform an ICCF analysis over the same observing interval, as shown in Appendix Figure \ref{fig:8}. The ICCF exhibits a clear positive peak, and the permutation test performed on its global maximum further indicates that this positive correlation is statistically significant. However, the correlation structure is relatively broad and contains multiple local peaks, and the ICCF results are also sensitive to interpolation. Therefore, the current data do not provide a robust constraint on a precise time lag. We thus refrain from further interpreting the detailed lag structure or inferring a causal sequence between the two quantities. Instead, we focus on the overall correlation behavior shown in Appendix Figure \ref{fig:8}, which indicates that the PRS emission and FRB activity exhibit coordinated variability on week-to-month timescales.

Overall, these results demonstrate that, on weekly to monthly timescales, the PRS flux density and the FRB burst energetics exhibit synchronous positive variability. A similar trend has also been mentioned in recent monitoring reports \citep{ATel}, consistent with the correlation identified in this work.

\subsection{FRB 20121102A}\label{c}
For FRB 20121102A, we focus our analysis on 2016, when both the PRS and FRB observations are relatively well-sampled. In later years, although FRB observations span a longer period (2020–2023), the corresponding PRS measurements are concentrated in 2022–2023, leaving limited temporal overlap between the two datasets (Figure \ref{fig:10}). This limits the meaningful comparison of the relative variability between the PRS and FRB activities. Consequently, the analysis below is primarily restricted to 2016. Figure \ref{fig:4} presents the S-band PRS flux density of FRB 20121102A from April to November 2016, alongside the corresponding daily FRB activity proxy. The related ICCF correlation analysis is shown in Figure \ref{fig:5}, where no significant correlation was found. Despite this, the PRS monitoring in 2016 still contains substantial temporal gaps, which may reduce the sensitivity of the correlation analysis. To mitigate this limitation, we further conduct the analysis over a more densely sampled sub-interval spanning August to October 2016. The ICCF results, presented in Figure \ref{fig:6}, also do not show any significant correlation.

\section{Discussion}

\subsection{Scintillation of the PRS Variability}
Propagation of radio waves from a point source through an inhomogeneous plasma can induce interstellar scintillation, which may account for the observed variability of the PRS \citep{1998MNRAS.294..307W}. The impact of scintillation on the PRS of FRB 20190520B has been previously investigated \citep{2023ApJ...959...89Z,2024ApJ...976..165Y}. These works suggest that the PRS of FRB 20190520B is subject to refractive scintillation, with a transition frequency of $\nu_{0}=12.5$ GHz along its line of sight. Following the formalism of \cite{1998MNRAS.294..307W}, the scattering strength is defined as $\xi=(\nu_0 / \nu)^{17/10}$, and the modulation index for refractive scintillation is given by $m_{\rm{p}}=\xi^{-1/3}$. At an observing frequency of 3 GHz, the expected modulation index is therefore $m_{\rm{p}}$ = 0.44. From the results in this paper, the $m_{\rm{obs}}=\sigma_S/\bar{S}=0.144$, where $\sigma_S$ and $\bar{S}$ are the standard deviation and mean of the PRS flux densities converted to 3 GHz, respectively, using the values listed in Table 4. Here, $m_{obs}$ characterizes the overall scatter of the adopted dataset within the corresponding observing window, rather than variability measured on a physical timescale. This value does not reach the estimated $m_p$. We further examine the possible contribution of refractive scintillation to the variability of the PRS associated with FRB 20240114A. The theoretically expected modulation index for this source at 3 GHz is $m_p \approx 0.484$ \citep{2025arXiv250114247Z}. From our measurements, we obtain an observed modulation index of
$m_{\rm obs} = \sigma_S / \bar{S} = 0.356$, where the corresponding values are calculated from the 3 GHz-converted PRS flux densities listed in Table 6. As above, $m_{obs}$ describes the overall scatter of the adopted dataset within the corresponding observing window. This value does not reach the predicted scintillation value. This suggests that, similar to the case of FRB 20190520B, the flux variability of the PRSs in both sources is more likely intrinsic in origin and may be physically connected to the same progenitor responsible for the FRB activity.

\subsection{PRS luminosity and FRB Energetics share a common Energy reservoir}
On weekly to monthly timescales, the PRS luminosity evolution shows a statistically significant positive correlation with the FRB burst energetics (Section \ref{a} and \ref{b}). This correlation suggests a shared energy reservoir between PRS and bursts. Within the magnetar central engine paradigm, the PRS may be powered by synchrotron emission from relativistic electrons \citep{Li_2020},
while the energy of FRBs is drawn from the same magnetic or rotational energy supply \citep{2020,Bochenek_2020}.

In the framework of a magnetar-powered MWN, the long-term evolution of the PRS luminosity is expected to gradually decline as the nebula expands and its energy density decreases. However, this secular evolution does not preclude variability on shorter timescales. On weekly to monthly timescales, fluctuations in the energy injection rate from the central engine or changes in particle acceleration conditions can lead to observable variations in the PRS flux density.

The bursts exhibit stochastic behavior, reflecting the sudden release of energy from the underlying reservoir. In contrast, the PRS, sustained by continuous injection of relativistic electrons from the magnetar, remains persistently observable. During active burst phases, the injection rate of relativistic electrons into the surrounding nebula may be temporarily enhanced, leading to an increase in PRS luminosity. If the injection spectrum hardens or the acceleration efficiency rises during these active phases, the resulting increase in synchrotron emissivity can manifest as an upturn in PRS luminosity.

The other repeating FRBs with PRSs have generally not shown a clear correlation between PRS luminosity and burst activity, likely because their burst-active windows are limited or sparsely sampled, restricting such analyses. FRB 20190520B is unique in providing both a PRS and sufficiently intensive burst monitoring to reveal a tentative connection between central engine activity and nebular emission. Similarly, FRB 20240114A also exhibits synchronous variability between the PRS luminosity and burst energetics. Although the overall monitoring duration is relatively short, the temporal overlap between PRS observations and burst activity measurements is comparatively high, enabling a meaningful correlation analysis. By contrast, FRB 20121102A does not exhibit a clear correlation between PRS luminosity and burst activity. This does not contradict a common physical origin. Rather, it indicates that while a detected correlation can provide evidence for a shared energy reservoir, its absence may result from additional modulation effects or differences in the evolutionary stage of the source. These comparisons highlight young or actively evolving systems such as FRB 20190520B and possibly FRB 20240114A as valuable laboratories for studying the interplay between PRS properties and FRB energetics.

Further progress requires more densely sampled PRS observations with improved temporal overlap with FRB activity. In particular, coordinated monitoring during different activity states (e.g., active versus quiescent phases) will be essential to better constrain the temporal relationship between PRS variability and burst energetics.

\section{CONCLUSION}
In this work, we systematically investigate several repeating FRB systems associated with PRSs. We find that for both FRB 20190520B and FRB 20240114A, the PRS flux density is positively correlated with the contemporaneous FRB burst-energy proxy on weekly to monthly timescales, with statistical significance supported by permutation-based tests. In contrast, other repeaters known to host PRSs have not yet shown compelling evidence for such a connection, likely due to limited burst monitoring, sparse temporal sampling, or burst activity confined to restricted active windows, all of which reduce the sensitivity of correlation analyses.

Our results suggest that the variability of PRS emission and FRB burst activity may be linked to a common central-engine energy reservoir, while being regulated by different physical processes. In this picture, the persistent radio emission traces the cumulative injection of relativistic particles into the surrounding nebula, whereas FRB bursts correspond to intermittent and impulsive energy-release events drawing from the same underlying energy supply.

We further find that the observed modulation indices of the PRSs are significantly lower than the theoretical expectations from refractive interstellar scintillation, indicating that the measured PRS variability cannot be explained by scintillation alone and likely contains a substantial intrinsic component.

Taken together, these results support a scenario in which the central engine of repeating FRBs continuously powers a surrounding synchrotron-emitting nebula while episodically producing FRB bursts. Continued long-term, multi-epoch monitoring of both burst energetics and PRS evolution will therefore be essential for clarifying the dynamical interplay between the central engine, the surrounding nebula, and the local environment.
%% Please use the acknowledgment and contribution environments. This will 
%% be anonomyized when the "anonymous" style option is used. 
\begin{acknowledgments}
We thank the FAST data center operations team for scheduling the follow-up observations and data acquisition. This work was supported by the National Natural Science Foundation of China (NSFC; Grant Nos. 12473039, 12588202, 12203069, 12503055, and 12563008) and partially supported by the New Cornerstone Science Foundation. C.-H. Niu also acknowledges support from the Basic Research Project of Central China Normal University, the Hubei QB Project, the CAS Youth Interdisciplinary Team and the Foundation of Guizhou Provincial Education Department (Grant No. KY(2023)059). J. R. Niu is supported by the Postdoctoral Fellowship Program of CPSF under Grant Number GZB20250737. Ju-Mei Yao was supported by the National Science Foundation of Xinjiang Uygur Autonomous Region (2022D01D85), the Major Science and Technology Program of Xinjiang Uygur Autonomous Region (2022A03013-2), and the CAS Project for Young Scientists in Basic Research (YSBR-063), the Tianshan talents program (2023TSYCTD0013), and the Chinese Academy of Sciences (CAS)"Light of West China"Program (No. xbzg-zdsys-202410 and No. 2022-XBQNXZ-015). R.S. Zhao thanks Science and Technology Foundation of Guizhou Provincial Department of Education (No. KY(2023)059)，Liupanshui Science and Technology Development Project (No. 52020-2024-PT-01).
\end{acknowledgments}

\bibliographystyle{aasjournalv7}
\bibliography{sample701}


\clearpage
\appendix

% Preamble:
% \usepackage{tabularx}
% \usepackage{booktabs}
% \usepackage{float}

\begin{table*}[h]
\centering
\normalsize
\setlength{\tabcolsep}{2.5pt}
\renewcommand{\arraystretch}{1.05}

\caption{\textbf{P$_{\rm day}$ Data of FRB 20190520B}}
\label{tab:2}

\noindent
% ===================== Column 1 =====================
\begin{minipage}[t]{0.24\textwidth}\centering
\vspace{0pt}
\begin{tabular}{@{}cc@{}}
\toprule
Date & $P_{\rm day}$ $^a$\\
\midrule
2020-07-21 & 0.5122 \\
2020-05-22 & 1.448 \\
2020-07-30 & 3.14 \\
2020-07-31 & 5.205 \\
2020-08-06 & 7.212 \\
2020-08-08 & 9.124 \\
2020-08-10 & 1.794 \\
2020-08-12 & 2.639 \\
2020-08-14 & 7.426 \\
2020-08-16 & 3.672 \\
2020-08-28 & 4.981 \\
2020-09-19 & 3.478 \\
2020-10-24 & 17.58 \\
2020-10-25 & 3.868 \\
2020-11-18 & - \\
2020-11-19 & - \\
2020-12-25 & 9.832 \\
2021-01-10 & 15.61 \\
2021-01-31 & - \\
2021-04-03 & 19.82 \\
2021-05-12 & 1.994 \\
2021-05-14 & 3.438 \\
2021-05-22 & 7.754 \\
2021-09-03 & - \\
2021-09-18 & - \\
2021-10-09 & - \\
2021-10-16 & - \\
2021-10-21 & - \\
2021-11-13 & 1.641 \\
2021-12-06 & 3.547 \\

\bottomrule
\end{tabular}
\end{minipage}
% ===================== Column 2 =====================
\begin{minipage}[t]{0.24\textwidth}\centering
\vspace{0pt}
\begin{tabular}{@{}cc@{}}
\toprule
Date & $P_{\rm day}$ \\
\midrule
2021-12-11 & 2.335 \\
2021-12-12 & 4.365 \\
2021-12-21 & 0.6887 \\
2021-12-22 & 3.516 \\
2021-12-25 & 5.867 \\
2021-12-27 & 0.8086 \\
2021-12-28 & 0.6266 \\
2021-12-29 & 1.847 \\
2021-12-30 & 1.689 \\
2021-12-31 & 0.8921 \\
2022-01-01 & 1.471 \\
2022-01-02 & 1.678 \\
2022-01-08 & - \\
2022-01-19 & 0.637 \\
2022-01-25 & - \\
2022-02-09 & 1.527 \\
2022-02-14 & 1.952 \\
2022-02-23 & 1.468 \\
2022-03-04 & 3.347 \\
2022-03-11 & 3.543 \\
2022-03-23 & 9.695 \\
2022-03-29 & 6.017 \\
2022-04-03 & 9.555 \\
2022-04-04 & 10.07 \\
2022-04-05 & 4.074 \\
2022-04-09 & 5.733 \\
2022-04-12 & 7.58 \\
2022-04-17 & 3.783 \\
2022-05-07 & 3.919 \\
2022-05-08 & 3.543 \\

\bottomrule
\end{tabular}
\end{minipage}
% ===================== Column 3 =====================
\begin{minipage}[t]{0.24\textwidth}\centering
\vspace{0pt}
\begin{tabular}{@{}cc@{}}
\toprule
Date & $P_{\rm day}$ \\
\midrule
2022-05-09 & 7.034 \\
2022-05-10 & 7.402 \\
2022-06-17 & 6.951 \\
2022-06-19 & 6.089 \\
2022-07-02 & 11.73 \\
2022-07-28 & 8.366 \\
2022-08-05 & 1.337 \\
2022-08-06 & 1.75 \\
2022-08-20 & 3.868 \\
2022-08-29 & - \\
2022-09-11 & 6.202 \\
2022-09-17 & 7.712 \\
2022-09-25 & 3.985 \\
2022-09-30 & 3.649\\
2022-10-01 & - \\
2022-10-19 & 9.037 \\
2022-11-06 & 4.528 \\
2022-11-11 & 6.873 \\
2022-11-27 & 32.41 \\
2022-12-24 & 12.62 \\
2023-01-16 & 3.693 \\
2023-02-05 & 6.608 \\
2023-02-13 & - \\
2023-02-16 & 3.927 \\
2023-03-02 & 3.674 \\
2023-04-01 & 0.3365 \\
2023-05-21 & 0.5204 \\
2023-05-25 & 0.653 \\
2023-05-26 & 0.4766 \\
2023-06-19 & 0.7229 \\

\bottomrule
\end{tabular}
\end{minipage}
% ===================== Column 4 =====================
\begin{minipage}[t]{0.24\textwidth}\centering
\vspace{0pt}
\begin{tabular}{@{}cc@{}}
\toprule
Date & $P_{\rm day}$ \\
\midrule
2023-07-07 & 1.192 \\
2023-08-17 & 0.6488 \\
2023-08-26 & - \\
2023-09-17 & - \\
2023-10-01 & 1.064 \\
2023-10-25 & 1.578 \\
2023-11-05 & 0.3934 \\
2023-11-06 & - \\
2023-11-12 & - \\
2023-12-10 & 1.979 \\
2023-12-27 & 1.064 \\
2024-01-18 & 3.144 \\
2024-03-12 & 1.517 \\
2024-03-28 & 1.38 \\
2024-04-03 & 2.723 \\
2024-04-13 & 1.53 \\
2024-04-15 & 1.8 \\
2024-04-25 & 2.189 \\
2024-04-27 & 1.173 \\
2024-04-29 & 3.149 \\
2024-05-01 & 4.432 \\
2024-05-13 & 1.657 \\
2024-05-16 & 4.136 \\
2024-06-05 & - \\
2024-06-16 & 6.892 \\
2024-07-17 & 17.73 \\
2024-07-31 & 33.73 \\
2024-09-03 & 7.316 \\
2024-10-26 & 3.198 \\
\bottomrule
\end{tabular}
\end{minipage}

\vspace{3mm}   % ← 控制与表格距离
\noindent
\begin{minipage}{\textwidth}
\footnotesize
\textbf{Notes.}\\
$^{a}$ The symbol – indicates that the noise diode was not injected for calibration on that observing day. The corresponding burst fluences are therefore unreliable and these data points are excluded from the correlation analysis.
\end{minipage}
\end{table*}

\begin{table*}[h]
\centering
\normalsize
\setlength{\tabcolsep}{2.5pt}
\renewcommand{\arraystretch}{1.05}

\caption{\textbf{The S-band–converted PRS flux densities of FRB 20190520B}}
\label{tab:3}

\noindent
% ===================== Column 1 =====================
\begin{minipage}[t]{0.45\textwidth}\centering
\vspace{0pt}
\begin{tabular}{@{}ccc@{}}
\toprule
Date & Telescope & Flux density ($\mu$Jy)\\
\midrule
2020-07-21 & VLA & 198.9±12.3 \\
2020-07-23 & VLA & 210.5±17.7\\
2020-08-17 & VLA & 182±12.6\\
2020-08-18 & VLA & 205.9±13.8\\
2020-08-29 & VLA & 198.3±11.3\\
2020-08-30 & VLA & 195±14\\
2020-09-12 & VLA & 177±8.53\\
2020-09-13 & VLA & 186±15\\
2020-09-15 & VLA & 192±12.6 \\
2020-09-19 & VLA & 176±16\\
2020-11-08 & VLA & 174.5±16.3\\
2020-11-14 & VLA & 233±17\\
2020-11-16 & VLA & 211±14\\
2021-10-01 & VLA & 148±19.2\\
2021-11-07 & VLA & 136±17.7\\
2022-02-26 & EVN & 159±32.3\\
2022-02-27 & EVN & 170±32.3\\
\bottomrule
\end{tabular}
\end{minipage}
% ===================== Column 2 =====================
\begin{minipage}[t]{0.45\textwidth}\centering
\vspace{0pt}
\begin{tabular}{@{}ccc@{}}
\toprule
Date & Telescope & Flux density ($\mu$Jy) \\
\midrule
2022-11-29 & GMRT &	196±32.2\\
2022-12-27 & GMRT &	114±19\\
2023-01-24 & GMRT &	205±30\\
2023-02-21 & GMRT &	87±26.5\\
2023-03-21 & GMRT &	205±18.3\\
2023-06-07  & VLA & 	157±9\\
2023-06-15  & VLA & 	173±11\\
2023-06-17	  & VLA & 171±12\\
2023-06-18 & VLA & 	166±7.4\\
2023-06-20	 & VLA & 147±4.76\\
2023-06-23 & VLA & 	147±13\\
2023-06-24 & VLA & 	148.7±9.8\\
2024-08-08 & VLA & 	170±10\\
2024-09-03 & VLA & 	168±17.5\\
2024-09-12 & VLA & 	134±11.7\\
2024-09-21 & VLA & 	139±14.6\\
\bottomrule
\end{tabular}
\end{minipage}
\end{table*}

\begin{table*}[h]
\centering
\normalsize
\setlength{\tabcolsep}{2.5pt}
\renewcommand{\arraystretch}{1.05}

\caption{\textbf{P$_{\rm day}$ Data of FRB 20240114A}}
\label{tab:6}

\noindent
% ===================== Column 1 =====================
\begin{minipage}[t]{0.24\textwidth}\centering
\vspace{0pt}
\begin{tabular}{@{}cc@{}}
\toprule
Date & $P_{\rm day}$ $^a$\\
\midrule
2024-06-16 & 36.626 \\
2024-06-23 &24.3684\\
2024-06-30 &21.6935\\
2024-07-07 &32.7147\\
2024-07-14 &114.124\\
\bottomrule
\end{tabular}
\end{minipage}
% ===================== Column 2 =====================
\begin{minipage}[t]{0.24\textwidth}\centering
\vspace{0pt}
\begin{tabular}{@{}cc@{}}
\toprule
Date & $P_{\rm day}$ \\
\midrule
2024-07-21& 173.398\\
2024-07-23 &97.9424\\
2024-07-25 &89.4095\\
2024-07-27 &148.509 \\
2024-07-28 &140.522\\
\bottomrule
\end{tabular}
\end{minipage}
% ===================== Column 3 =====================
\begin{minipage}[t]{0.24\textwidth}\centering
\vspace{0pt}
\begin{tabular}{@{}cc@{}}
\toprule
Date & $P_{\rm day}$ \\
\midrule
2024-07-29& 199.112\\
2024-07-31 &114.669\\
2024-08-03& 121.384\\
2024-08-06 &40.6069\\
2024-08-09  &81.004\\
\bottomrule
\end{tabular}
\end{minipage}
% ===================== Column 4 =====================
\begin{minipage}[t]{0.24\textwidth}\centering
\vspace{0pt}
\begin{tabular}{@{}cc@{}}
\toprule
Date & $P_{\rm day}$ \\
\midrule
2024-08-15& 125.8\\
2024-08-18& 82.8507\\
2024-08-22 & 79.541\\
2024-08-30 &37.7285\\

\bottomrule
\end{tabular}
\end{minipage}
\end{table*}

\begin{table*}[h]
\centering
\normalsize
\setlength{\tabcolsep}{2.5pt}
\renewcommand{\arraystretch}{1.05}

\caption{\textbf{The S-band–converted PRS flux densities of FRB 20240114A}}
\label{tab:7}

\noindent
% ===================== Column 1 =====================
\begin{minipage}[t]{0.45\textwidth}\centering
\vspace{0pt}
\begin{tabular}{@{}ccc@{}}
\toprule
Date & Telescope & Flux density ($\mu$Jy)\\
\midrule
2024-06-15 & GMRT & 36.0±9.3 \\
2024-07-03 & GMRT & 30.4±6.9\\
2024-07-07 & GMRT & 37.4±11.6\\
2024-07-13 & GMRT & 53.9±14.4\\
2024-07-23 & VLA & 97.2±11.9\\
2024-07-27 & VLA & 79.5±6.5\\
\bottomrule
\end{tabular}
\end{minipage}
% ===================== Column 2 =====================
\begin{minipage}[t]{0.45\textwidth}\centering
\vspace{0pt}
\begin{tabular}{@{}ccc@{}}
\toprule
Date & Telescope & Flux density ($\mu$Jy) \\
\midrule
2024-07-28 & GMRT &	44.5±11.6\\
2024-07-29 & VLA &	84.8±6.9\\
2024-08-22 & GMRT &	42.0±9.2\\
2024-09-28 & VLBA &	54.7±10.7\\
2024-09-30 & VLBA &	54.7±10.7\\
2024-10-03 & VLA & 	66.8±7.2\\
\bottomrule
\end{tabular}
\end{minipage}

\end{table*}

\begin{figure*}[h]
    \centering
    \includegraphics[width=0.9\linewidth]{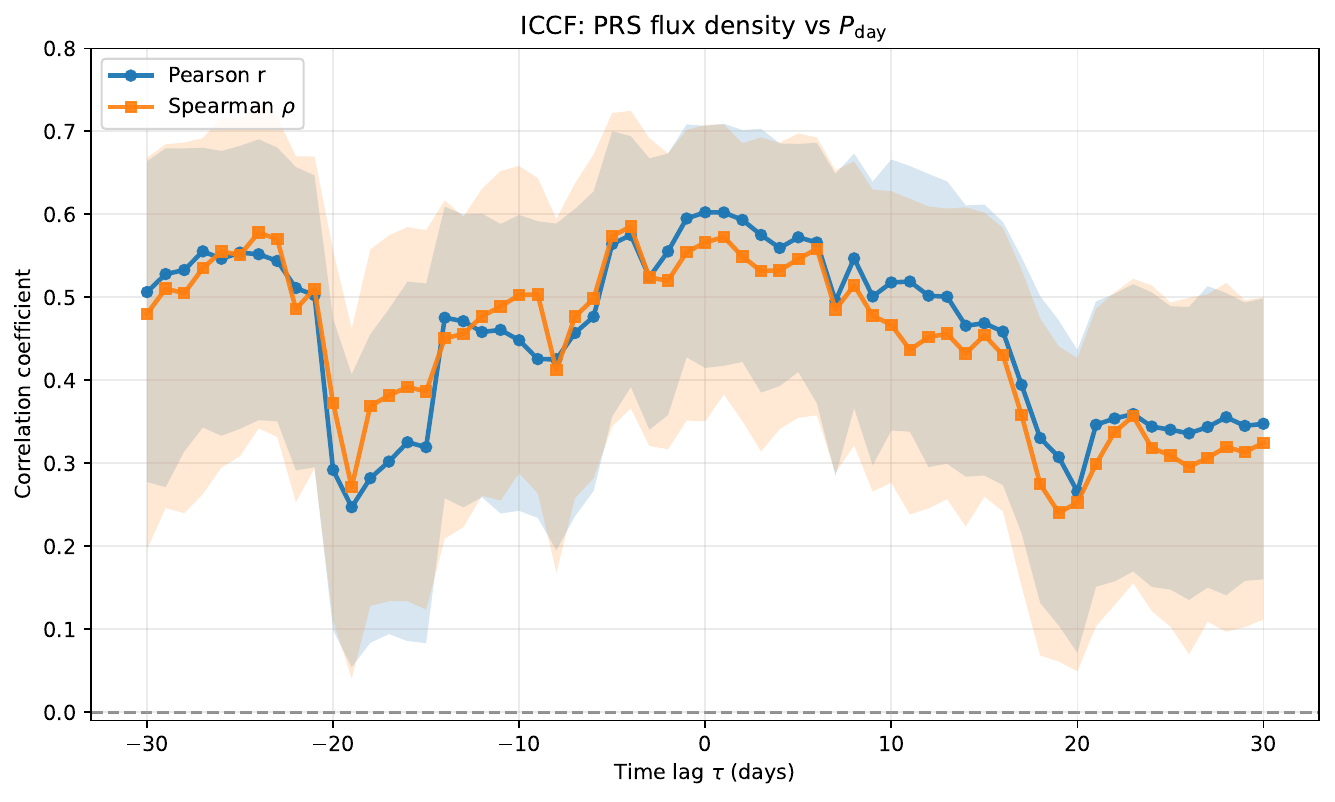}
    \includegraphics[width=0.9\linewidth]{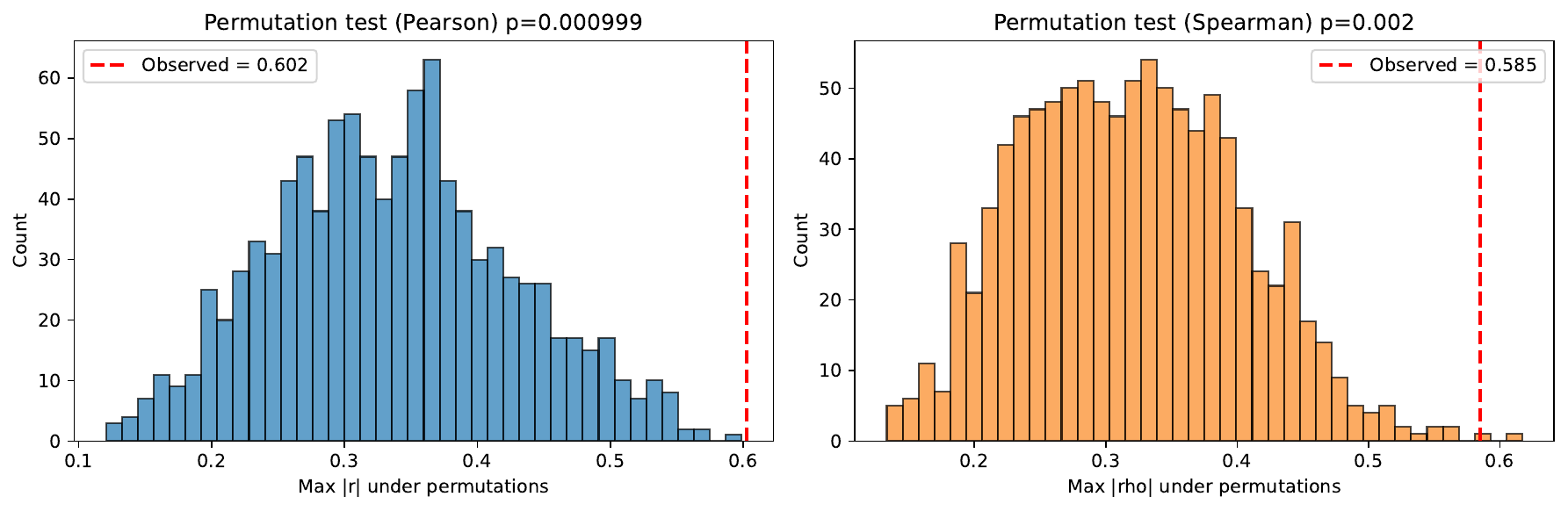}
    \caption{\textbf{Permutation-test results for the June 2020--June 2022 data segment of FRB 20190520B based on the ICCF analysis. }}
    \label{fig:2}
\end{figure*}

\begin{figure*}[h]
    \centering
    \includegraphics[width=0.9\linewidth]{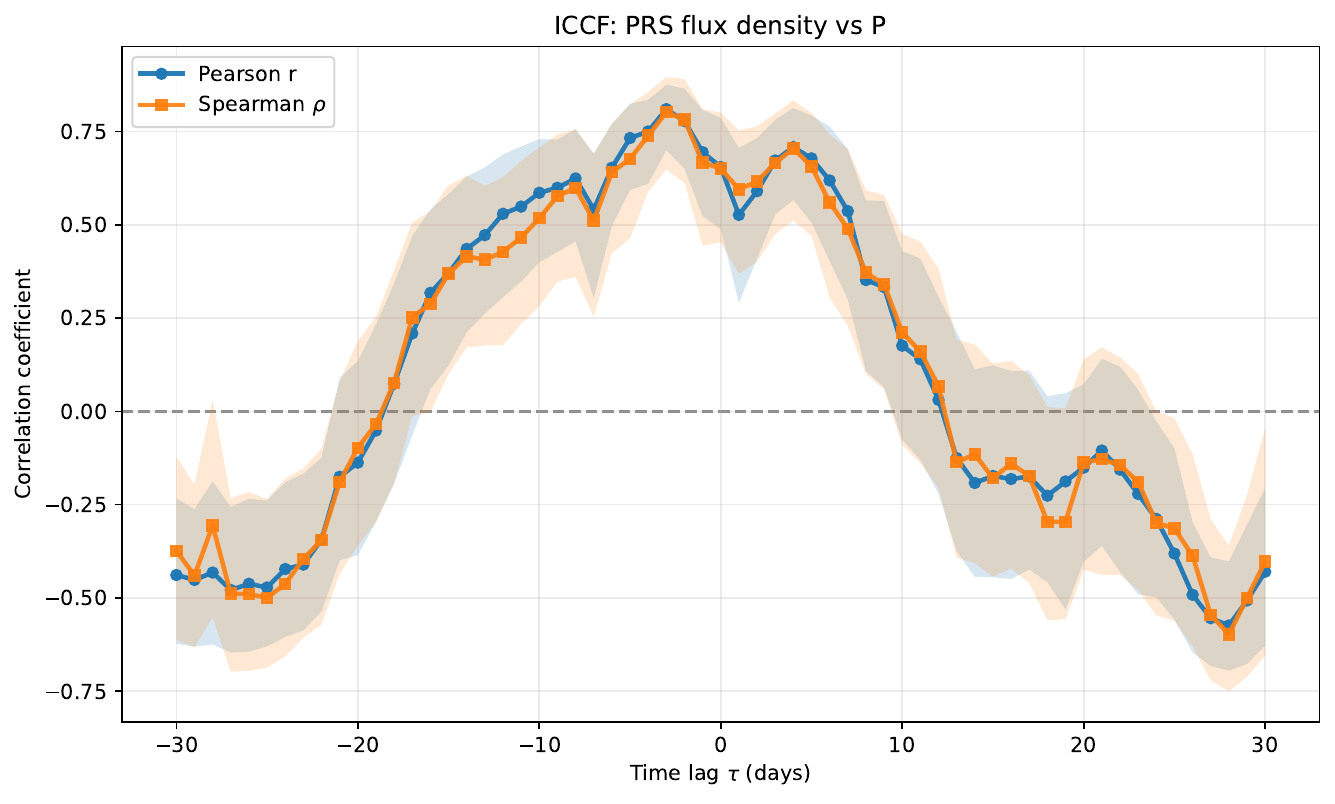}
    \includegraphics[width=0.9\linewidth]{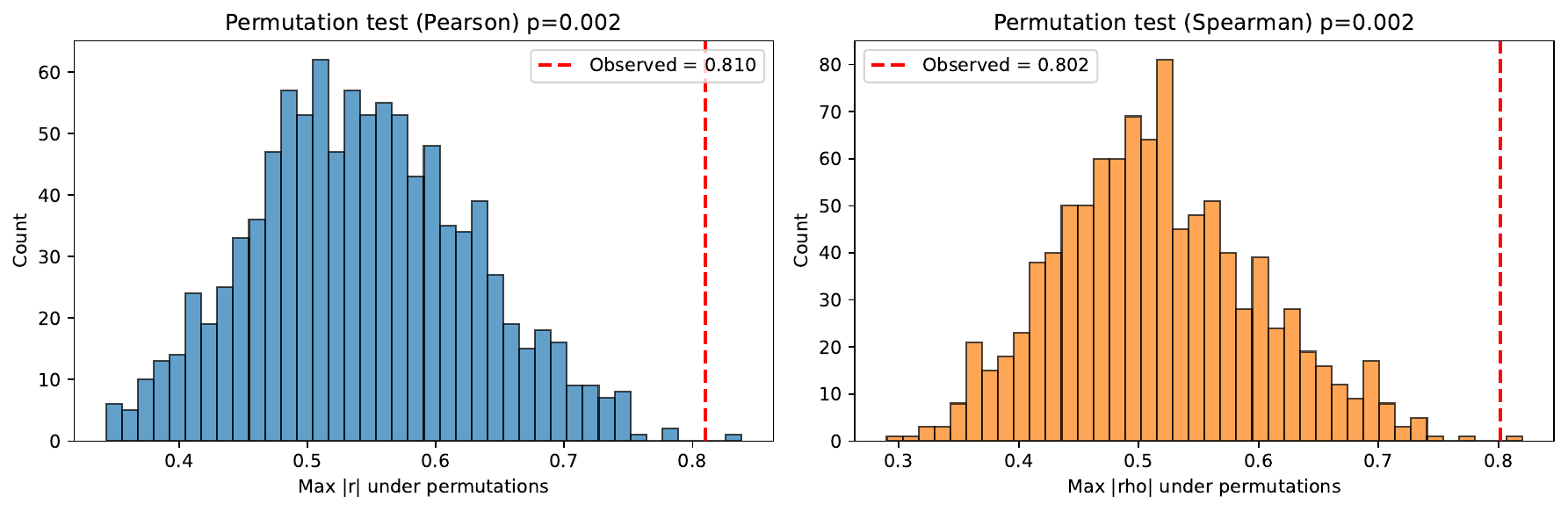}
    \caption{\textbf{Permutation-test results for FRB 20240114A based on the ICCF analysis. }}
    \label{fig:8}
\end{figure*}

\begin{figure*}[h]
    \centering
    \includegraphics[width=0.8\linewidth]{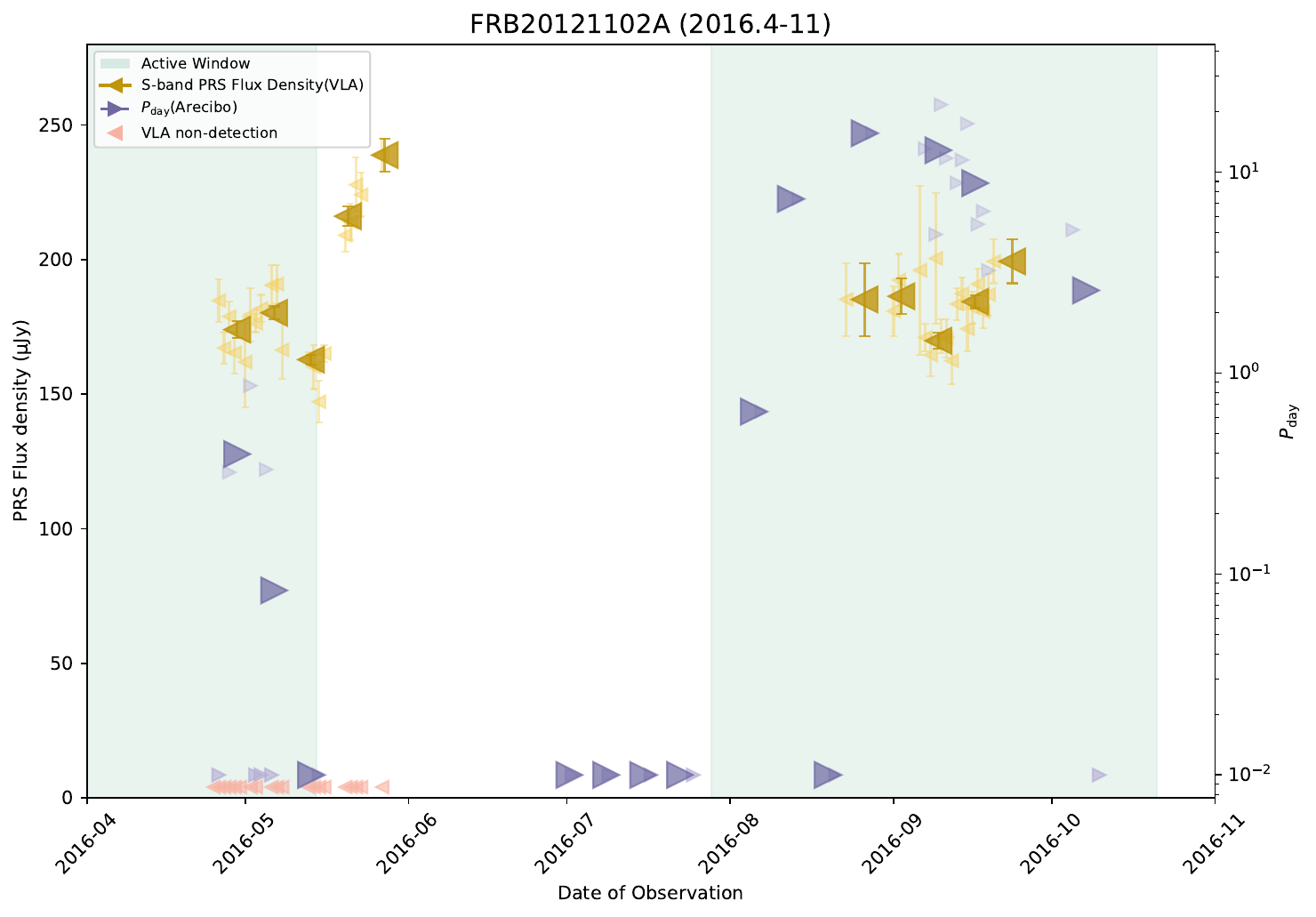}
    \caption{\textbf{S-band PRS flux density of FRB 20121102A and the corresponding daily FRB activity proxy, $P_{\rm day}$}. The symbol definitions are the same as those used for FRB 20190520B. Unlike FRB 20190520B, however, the S-band PRS flux densities shown here are direct measurements and are not converted from other frequencies using a spectral index. The red markers along the bottom indicate VLA burst-search epochs in May 2016, during which no FRB bursts were detected. Given the limited number of data points and the relatively short temporal coverage, the binned values are computed using a 7-day binning scheme starting from the first PRS observation. Previous studies have shown that FRB 20121102A exhibits activity cycles in the L band, with a period of approximately 159 days \citep{2025A&A...693A..40B,2026AA...707A.335E}, of which about 84 days correspond to an active phase. The light-green shaded regions indicate the time intervals during which the source is expected to be in a relatively active state.}
    \label{fig:4}
\end{figure*}

\begin{figure}[h]
    \centering
    \includegraphics[width=0.9\linewidth]{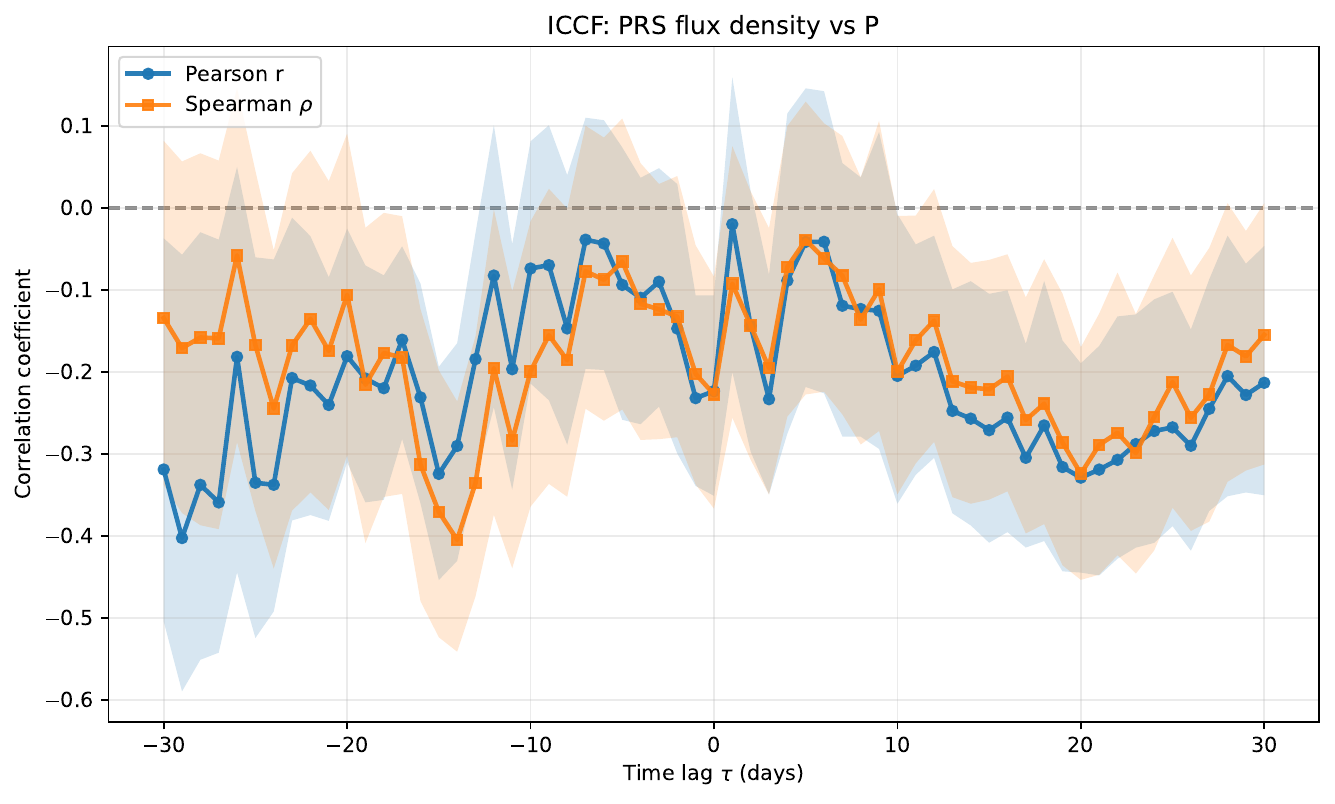}
    \includegraphics[width=0.9\linewidth]{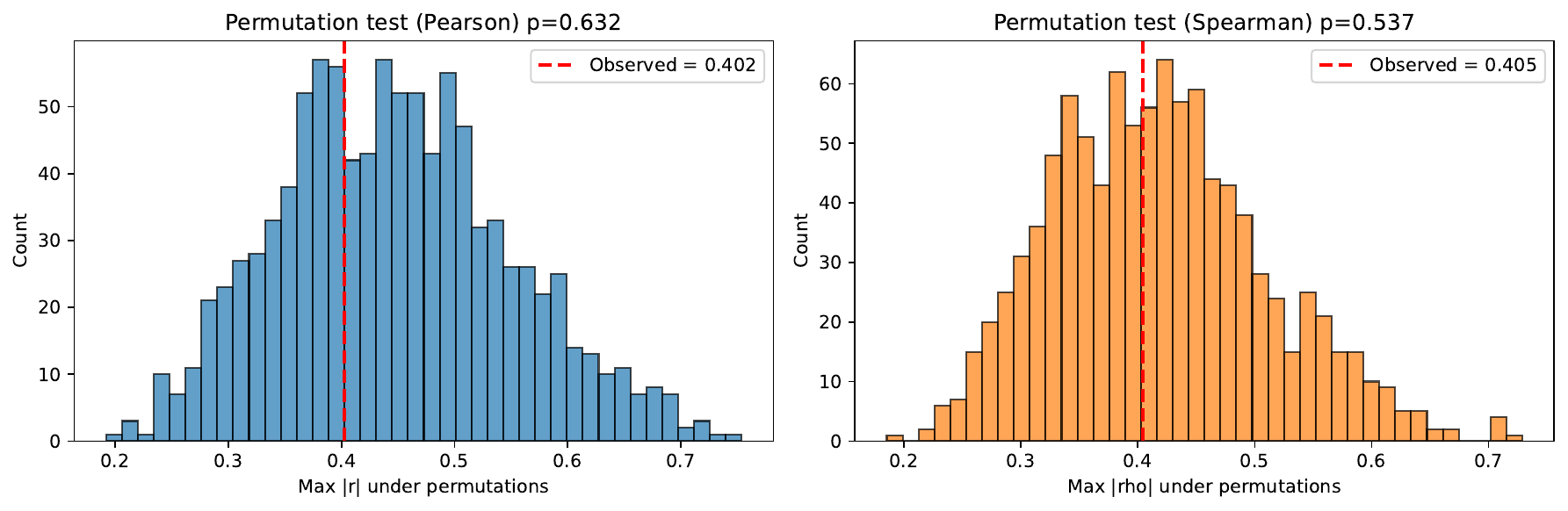}
    \caption{\textbf{Permutation-test results for FRB 20121102A based on the ICCF analysis over the 2016 April–November interval.}}
    \label{fig:5}
\end{figure}

\begin{figure}
    \centering
    \includegraphics[width=0.9\linewidth]{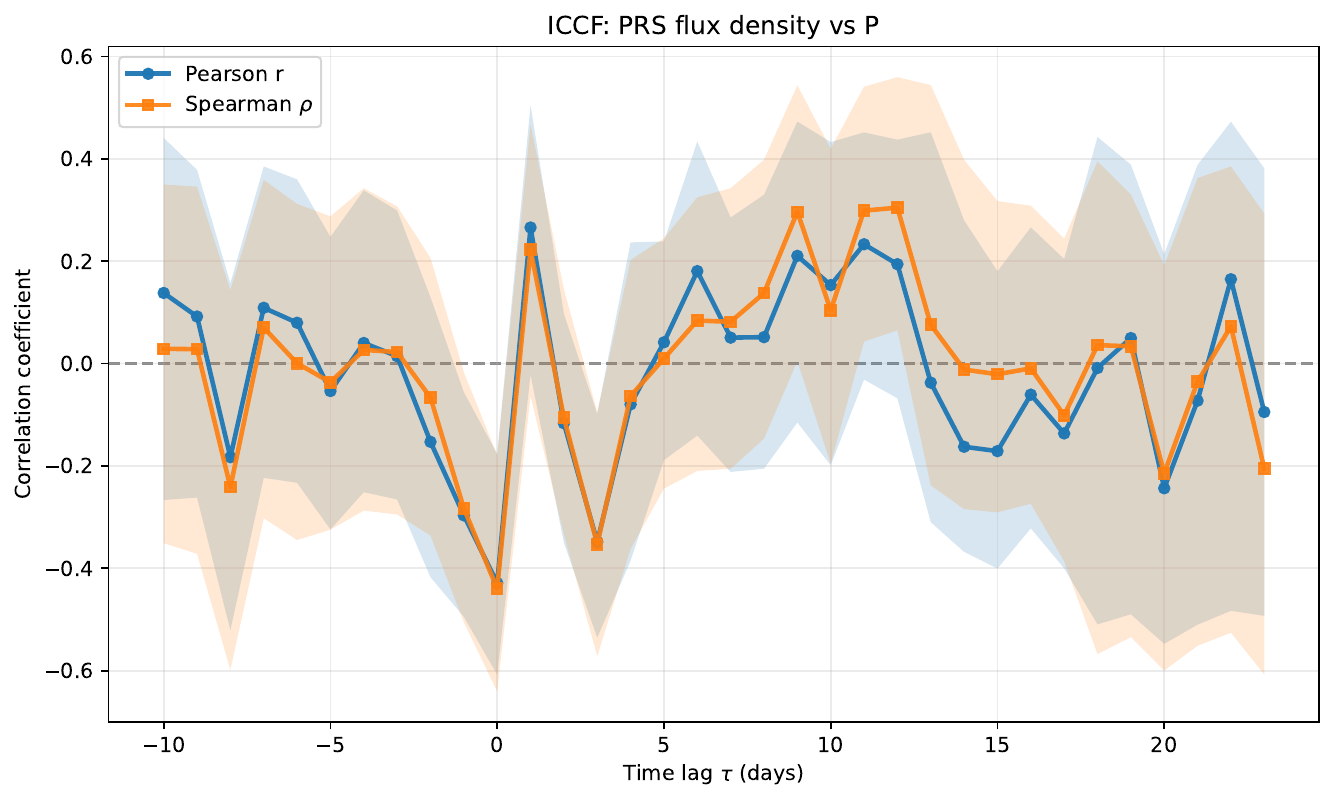}
    \includegraphics[width=0.9\linewidth]{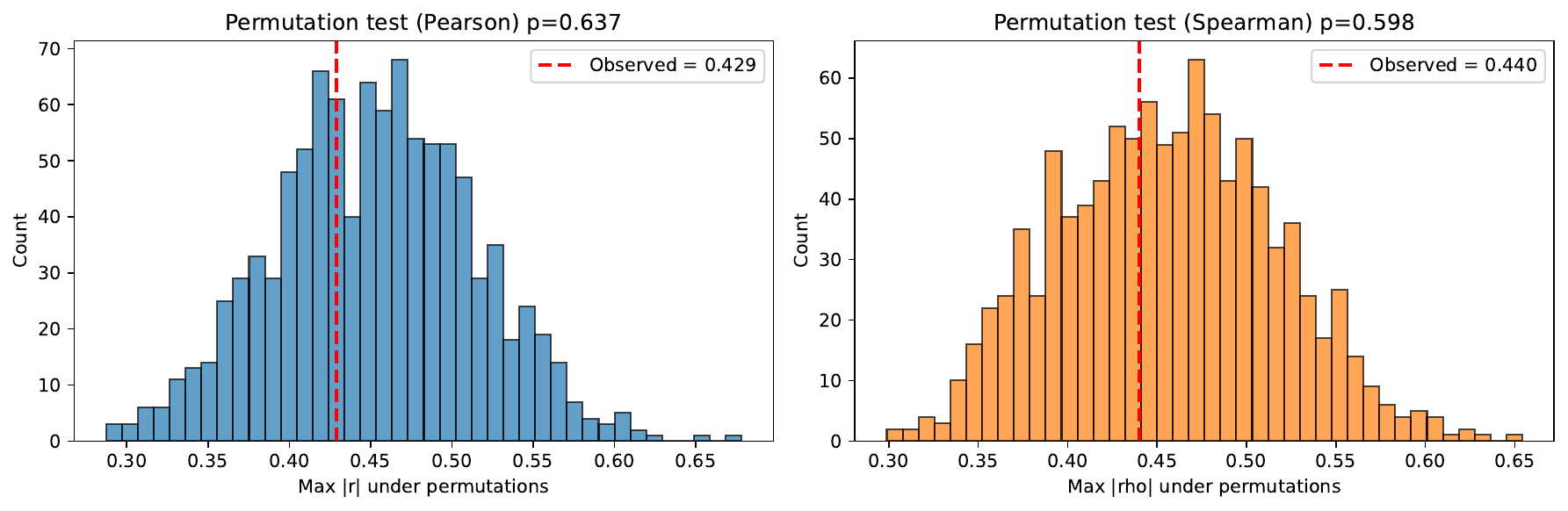}
    \caption{\textbf{Permutation-test results for FRB 20121102A based on the ICCF analysis over the 2016 August–October sub-epoch.}Correlation coefficients are reported only for lags where the number of valid paired points after interpolation exceeds the adopted minimum threshold; lags with insufficient effective pairs are left blank.}
    \label{fig:6}
\end{figure}

\begin{figure}
    \centering
    \includegraphics[width=0.8\linewidth]{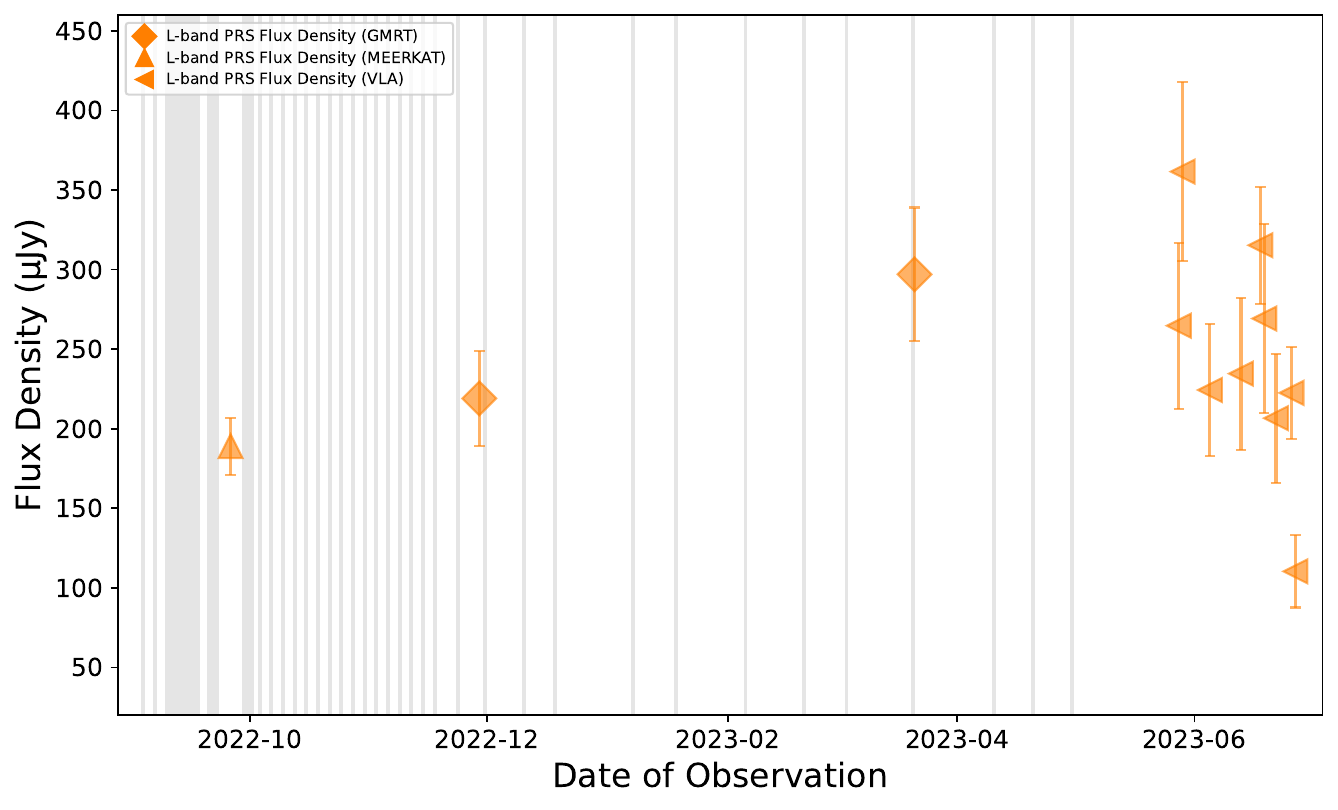}
    \caption{PRS flux-density measurements of FRB 20121102A in 2022–2023, with gray shaded intervals indicating the epochs of FAST monitoring.}
    \label{fig:10}
\end{figure}

\begin{figure}
    \centering
    \includegraphics[width=0.8\linewidth]{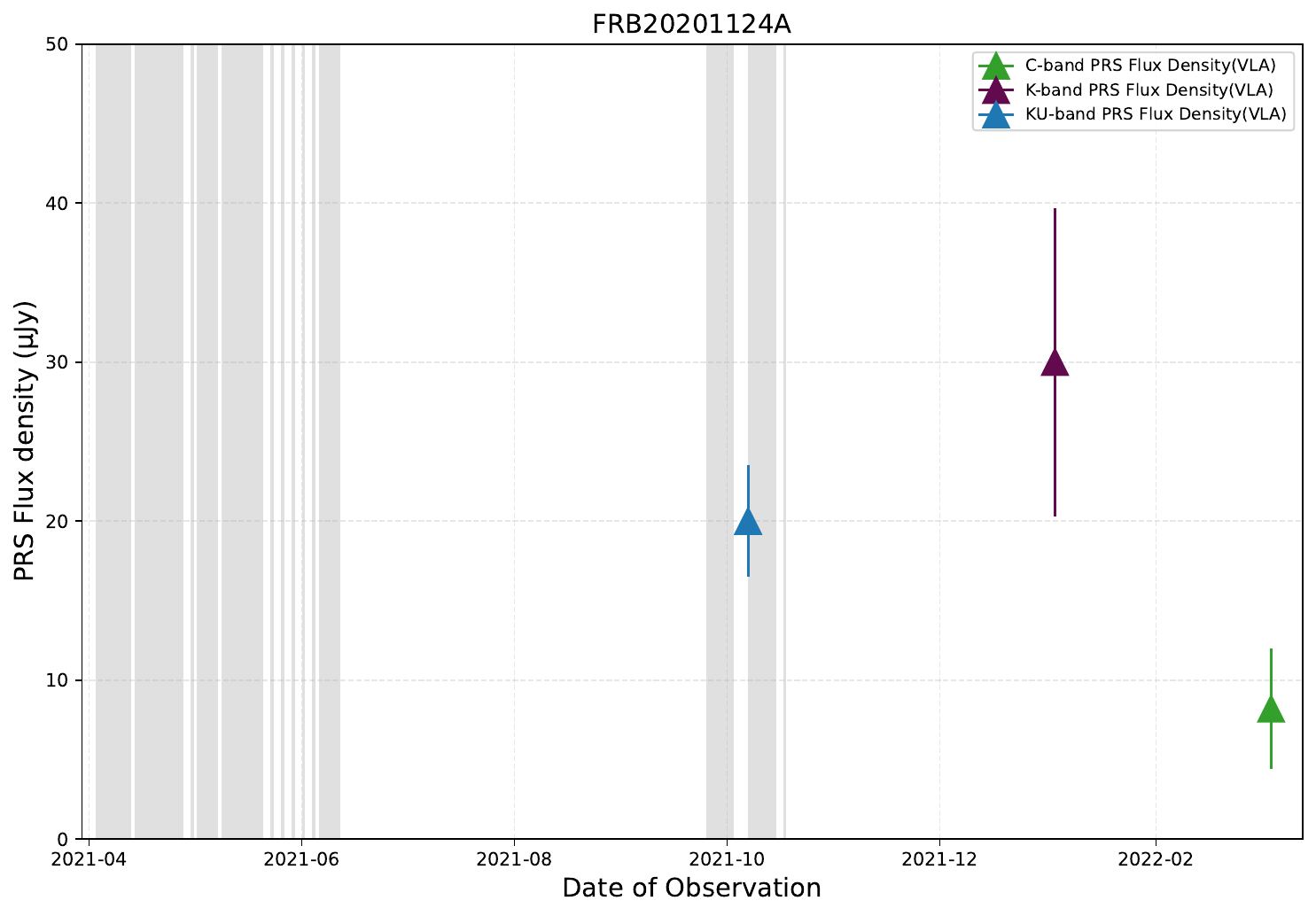}
    \caption{PRS flux-density measurements of FRB 20201124A, with gray shaded intervals indicating the epochs of FAST monitoring.}
    \label{fig:7}
\end{figure}

\end{document}